\documentclass[aps, groupedaddress, amsmath, amssymb, floatfix, twocolumn, pre]{revtex4-1}
\usepackage{amsmath,amssymb,amsfonts,bm,cancel}
\usepackage{graphicx}
\usepackage{tabularx}
\usepackage{url}
\usepackage{color,xcolor,soul}
\setstcolor{red}

\usepackage[columnwise, modulo]{lineno}

\begin{document}

\title{Covid-19 epidemic under the K-quarantine model: Network approach}

\author{K. Choi}
\affiliation{CCSS, CTP and Department of Physics and Astronomy, Seoul National University, Seoul 08826, Korea}
\author{Hoyun Choi}
\affiliation{CCSS, CTP and Department of Physics and Astronomy, Seoul National University, Seoul 08826, Korea}
\author{B. Kahng}
\email{bkahng@snu.ac.kr}
\affiliation{CCSS, CTP and Department of Physics and Astronomy, Seoul National University, Seoul 08826, Korea}

\date{\today}

\begin{abstract}
The Covid-19 pandemic is ongoing worldwide, and the damage it has caused is unprecedented. For prevention, South Korea has adopted a local quarantine strategy rather than a global lockdown. This approach not only minimizes economic damage, but it also efficiently prevents the spread of the disease. In this work, the spread of COVID-19 under local quarantine measures is modeled using the Susceptible-Exposed-Infected-Recovered model on complex networks. In this network approach, the links connected to isolated people are disconnected and then reinstated when they are released. This link dynamics leads to time-dependent reproduction number. Numerical simulations are performed on networks with reaction rates estimated from empirical data. The temporal pattern of the cumulative number of confirmed cases is then reproduced. The results show that a large number of asymptomatic infected patients are detected as they are quarantined together with infected patients. Additionally, possible consequences of the breakdowns of local quarantine measures and social distancing are considered.
\end{abstract}
\maketitle

\section{Introduction}
The COVID-19 pandemic has changed various aspects of our societies, ranging from public health and economic conditions to human rights.
Two other recent coronavirus pandemics, Severe Acute Respiratory Syndrome (SARS) in 2002 and Middle East Respiratory Syndrome (MERS) in 2013, have produced 8437 and 2519 cases, respectively~\cite{SARS, MERS}. On the other hand, within just eight months (as of September 4th, 2020) there have been about 28 million cases of Covid-19 and 0.9 million resulting deaths ~\cite{COVID}.
This is due to an abnormally high transmission rate, asymptomatic spreading, and the lack of vaccines or treatments~\cite{CDC, Liu2020}.
Under these circumstances, non-pharmaceutical interventions such as social distancing among individuals, masking, and reinforcing personal hygiene are alternative approaches to prevention.

Beginning with Wuhan~\cite{Wu2020-1,Wu2020-2}, China, the majority of countries facing the spread of COVID-19 have used the lockdown policy that restricts travel from other countries and prevents people from participating in non-essential social activities~\cite{reka, vespignani-global1,vespignani-global2,Hsiang2020,Giordano2020, secondWave}. However, such a lockdown policy is not sustainable, because it drastically reduces economic activities~\cite{reka}. Indeed, the majority of countries that adopted the lockdown policy have failed to sustain it for more than two to three months; they are gradually returning to their former policies.

The Korean Center for Disease Control and Prevention (KCDC) has achieved great success using the so-called K-quarantine model, which enforces local quarantine around confirmed patients rather than implementing a global lockdown. This approach, implemented using the ``3T steps,'' efficiently prevents the spread of disease without critical economic damage.

The first step in this procedure is testing. In South Korea, based on the measures adopted during the 2015 MERS outbreak, diagnostic tool kits have been developed. Thus, the Covid-19 inspection capability and speed have improved drastically. Moreover, using the drive-through and walk-through methods, rapid and large-scale testing with safe separation between potential patients and medical staff has become possible. Individuals who enter the country from abroad or have been in close contact with confirmed patients are required undergo diagnostic tests. Those with positive results are required to be placed in quarantine for two weeks, which is the maximum incubation period of the disease. During the quarantine period, they must monitor their health daily and report it on a self-quarantine safety protection mobile application managed by the KCDC. Their locations are also monitored using the application. If they exit officially designated locations, they are immediately apprehended and fined. Thus, a near-ideal quarantine system is achieved.

The second step is tracing. When the diagnostic test confirms that an individual is Covid-19 positive, an epidemiological investigation system is launched for that patient. The KCDC traces every place the patient visited at any time over a recent time window. These places and times are immediately reported to the public (without any personal information of the patient) using the application within the districts of visited sites. Every individual that the patient has been in contact with is identified and requested to take a diagnostic test. Those who were in close contact with the patient are preemptively quarantined, while others are encouraged to self-isolate at home. Among them, if some cases are confirmed, the infection route is identified. Thus, the locations at which the epidemic has originated are found, closed, and disinfected. Those who were in the same place but did not have close contact with the patient are identified or encouraged to voluntarily report to the KCDC, which allows them to take free diagnostic tests.
In certain cases, the location identified may involve privacy infringement, such that individuals present at the same time as the patient may hesitate to report themselves having been there. Such individuals are identified using resources such as the mobile phone records in the local station and required to take a diagnostic test. Further, they are requested to self-quarantine. During the quarantine period, individuals are also expected to check their body temperature and report it on the application. After two weeks, they are encouraged to undergo the diagnostic test again. If the test result is negative, they are released and can return to their normal lives.

The final step is treatment. For efficient usage of medical resources, the severity of patients' prompts are graded. Serious cases are admitted to either the residential treatment center or the hospital. Others are isolated at low-level designated locations or at home. The distribution of medical resources and facilities has contributed significantly to the low death rate in South Korea.
In addition, hospitals isolate respiratory patients to protect ordinary patients who require urgent medical care and are availing other medical services.

These K-quarantine strategies may have some side effects such as invasion of privacy and infringement of human rights. Therefore, the KCDC makes a significant effort to conceal patients' personal information from the public. 

\begin{figure*}[!t]
	\centering
	\includegraphics[width=0.95\linewidth]{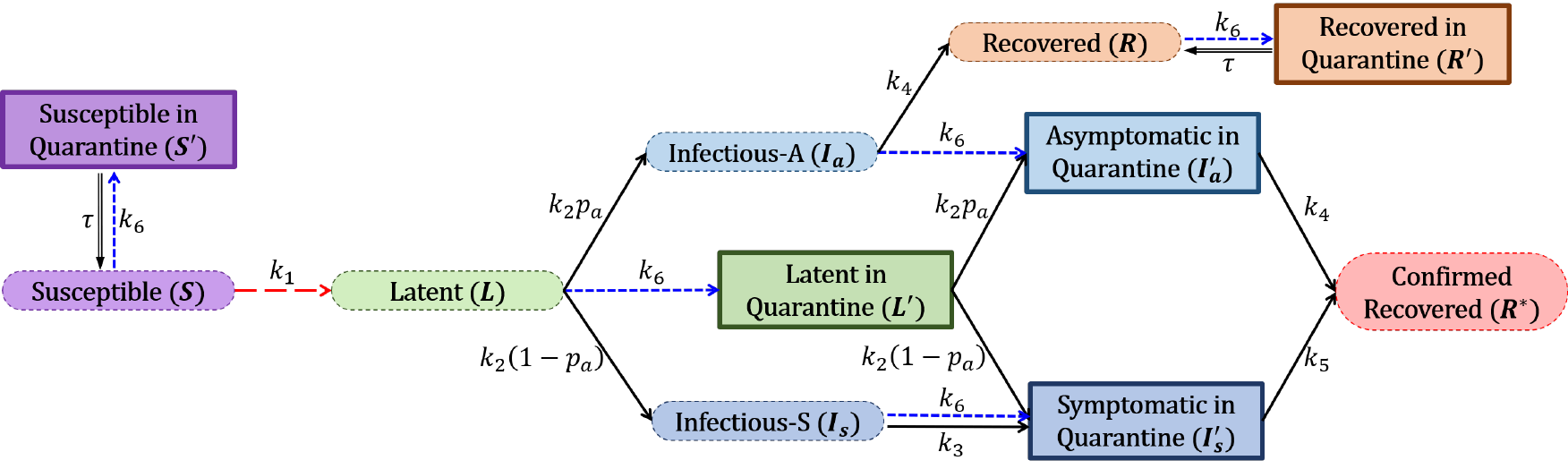}
	\caption{
	Flowchart for the K-quarantine model.
	The states under quarantine are represented by squares, while others are represented by stadiums.
	\label{fig:diagram}
	}
\end{figure*}

\section{Networks}
It is worth examining the manner in which the epidemic contagion spreads under the K-quarantine model as compared to its spread under global lockdown without local restrictions.
To achieve this, a mathematical model is considered in this work.
The conventional epidemiological model is a compartmental model in which each person is considered to be in one of the following possible states: susceptible ($S$), latent ($L$), infected ($I$), or recovered/deceased ($R$).
The proportions of people in each state are regarded as continuous variables, and their rate equations (time derivatives) are set up as a function of these proportions with appropriate rate constants.
By solving these differential equations, the fraction of each state as a function of time is obtained.
In the past, this approach has successfully predicted the evolution of the fraction of infected populations.
However, it may not be useful when considering the local quarantine effect under the K-quarantine measures stated above.

Here, the epidemic reactions are simulated on networks. A network is composed of nodes and links, which represent people and contact between a pair of connected people, respectively. The numbers of nodes and links that are simulated on are taken as $N = 2.1 \times 10^4$ and $L = 5 \times N$, respectively. This implies that a society composed of $N$ people is being considered, and the average number of people in contact with each person (called the `degree' in graph theory) is given as $\langle d \rangle= 2L/N = 10$. Some of them, such as family members and colleagues in the workplace, are in close contact, whereas others, such as people who met at shopping malls, are in loose contact. However, these two groups were not differentiated in the simulations described herein. This is because distinguishing between these two types of contact is only necessary for lockdown measures~\cite{reka}, where the loosely connected links are disconnected. In the K-quarantine model, such a lockdown is not applied. The links need not be distinguished into two types. Instead, all links are regarded as close contacts in this small simulation system sizes. The K-quarantine process is realized by locally disconnecting the links to an infected node. As soon as the quarantine is completed and the patient is released, these links are reinstated. In the K-quarantine model, once a person is quarantined, they are required to take a diagnostic test. If the result is positive, then the people in contact with the patient are quarantined. Thus, links connected to the neighbors of the confirmed patient also need to be disconnected.

Networks are classified into two types based on their connection configurations: random networks and scale-free networks.
For random networks, each link is added between two randomly selected nodes. Thus, the degrees of each node have a Poisson distribution. Because this model was first proposed by Erd\H{o}s–R\'{e}nyi, it is often called the ER model~\cite{Erdos1960}. For scale-free networks, following the power law, the degrees of each node are heterogeneous. This implies that a few nodes have large degrees, but the remaining nodes have small ones. The nodes with large numbers of neighbors are called hubs. When a hub is infected, a large number of susceptible neighbors are exposed to the contagion. This may result in a spike in contagion. Scale-free networks were constructed using the models proposed by Goh et al.~\cite{Goh2001} and Chung and Lu~\cite{Chung2002}.

\section{Models}
The epidemic reactions proceed as per Markovian dynamics, which are realized by the Gillespie algorithm (GA)~\cite{Gillespie1977,Vestergaard2015}.
Each node is in one of the following states~\cite{Chen2020,Davies2020,secondWave,He2020,Pei2009}: susceptible ($S$), latent ($L$), asymptomatic infectious ($I_a$), symptomatic infectious ($I_s$), asymptomatic in quarantine($I_a^\prime$), symptomatic in quarantine ($I_s^\prime$), or recovered ($R$, $R^\prime$, or $R^*$)~\cite{asymp_presymp, presymp1, presymp2, contactTrace}.
The states of susceptible in quarantine ($S^\prime$) and latent in quarantine ($L^\prime$) also exist. The dynamic begins with one infected person, with all the others being in a susceptible state. When nodes in states $L$, $I_a$, and $I_s$ are absent, the dynamic falls into an absorbing state, and the nodes in state $S$ or $R$ remain. The detailed dynamics are as follows:

A susceptible individual in contact with an infectious individual $I_a$ and $I_s$  enters the latent state ($L$) at the rate $k_1$.
These reactions are expressed as
\begin{linenomath*}\begin{align}\label{eq:S2E}
	S + I_a \xrightarrow{k_{1}} L + I_a \quad \textrm{and} \quad  S + I_s \xrightarrow{k_{1}} L + I_s.
\end{align}\end{linenomath*}
When the latency period ends, the individual becomes infectious, that is, they can transmit the infection with or without symptoms. These states are denoted as $I_a$ or $I_s$, respectively. These processes occur at rates $k_2 p_a$ and $k_2 (1-p_a)$, respectively. Here, $p_a$ represents the fraction of asymptomatic infectious patients. These reactions are expressed as
\begin{linenomath*}\begin{align}\label{eq:E2AI}
	L \xrightarrow{k_{2}p_a} I_a \quad \textrm{and} \quad L \xrightarrow{k_2(1-p_a)} I_s.
\end{align}\end{linenomath*}

When symptoms develop, the infected individual must go to the hospital and take a diagnostic test.
If the result is positive, they are quarantined.
This process occurs at the rate $k_3$ and is expressed as
\begin{linenomath*}\begin{align}\label{eq:I2QI}
	I_s \xrightarrow{k_{3}} I_s^{\prime},
\end{align}\end{linenomath*}
where the prime indicates that the individual is quarantined.
On the other hand, an asymptomatic individual may recover naturally without any treatment.
This process occurs at the rate $k_4$ and is expressed as
\begin{linenomath*}\begin{align}\label{eq:A2R}
	I_a \xrightarrow{k_4} R.
\end{align}\end{linenomath*}
The isolated individual in state $I_s^{\prime}$ may be recovered through treatment or succumb to the disease.
This recovered individual is counted as a confirmed case of recovery, denoted by $R^{*}$.
This process occurs at the rate $k_5$ and is expressed as
\begin{linenomath*}\begin{align}\label{eq:QI2CR}
	I_s^{\prime} \xrightarrow{k_5} R^{*}.
\end{align}\end{linenomath*}

In the K-quarantine model, confirmed cases ($I_s^{\prime}$) and their neighbors are self-quarantined as potential infectious people even if they are asymptomatic.
Regardless of their state being $S$, $L$, $I_a$, $I_s$, or $R$, they are quarantined at the rate $k_6$.
This process is expressed as
\begin{linenomath*}\begin{align}\label{eq:X2QX}
	I_s^{\prime} + X \xrightarrow{k_{6}} I_s^{\prime} + X^{\prime},\quad X \in \{S,E,I_a,I_s,R \},
\end{align}\end{linenomath*}
where $k_{6}$ is the quarantine rate.
Because quarantined individuals must undergo a diagnostic test, isolated asymptomatic carriers $I_a^\prime$ are identified as confirmed cases.
Accordingly, the neighbors of the identified asymptotic carrier are also quarantined at the rate $k_6$:
\begin{linenomath*}\begin{align}\label{eq:X2QX_2}
	I_a^{\prime} + X \xrightarrow{k_{6}} I_a^{\prime} + X^{\prime},\quad X \in \{S,E,I_a,I_s,R \},
\end{align}\end{linenomath*}
This trace process is repeated until no further confirmed cases are identified.~\cite{contactTrace, contactTrace2}.
During the quarantine period, identified asymptomatic infected individuals recover at the rate $k_4$, expressed as
\begin{linenomath*}\begin{align}\label{eq:QA2CR}
	I_a^{\prime} \xrightarrow{k_{4}} R^{*}.
\end{align}\end{linenomath*}
Here, it is assumed that asymptomatic patients have the same recovery rate $k_4$ regardless of isolation.
Individuals in the states $S^{\prime}$, $L^{\prime}$, or $R^{\prime}$ with negative diagnostic test results are released from quarantine. They then return to their original states.
\begin{linenomath*}\begin{align}\label{eq:QX2X}
	X^{\prime} \stackrel{\tau}{\Rightarrow} X,\quad X \in \{S,L,R\},
\end{align}\end{linenomath*}
where $\tau$ is the quarantine period (not the rate).

The reproduction number (denoted as $R_0$), the number of individuals who are susceptible and become infectious by contacting an infected individual, is calculated as $R_0 = k_1 \langle d \rangle /k_3$, where $\langle d \rangle$ is the mean number of neighbors on a given network. Herd immunity is the level of immunity in a population that prevents the spread of a disease over the entire system. The herd immunity threshold is described as $P_c=1-(1/R_0)$~\cite{herd,Gani2005}.

\begin{figure*}[!th]
\centering
\includegraphics[width=0.815\linewidth]{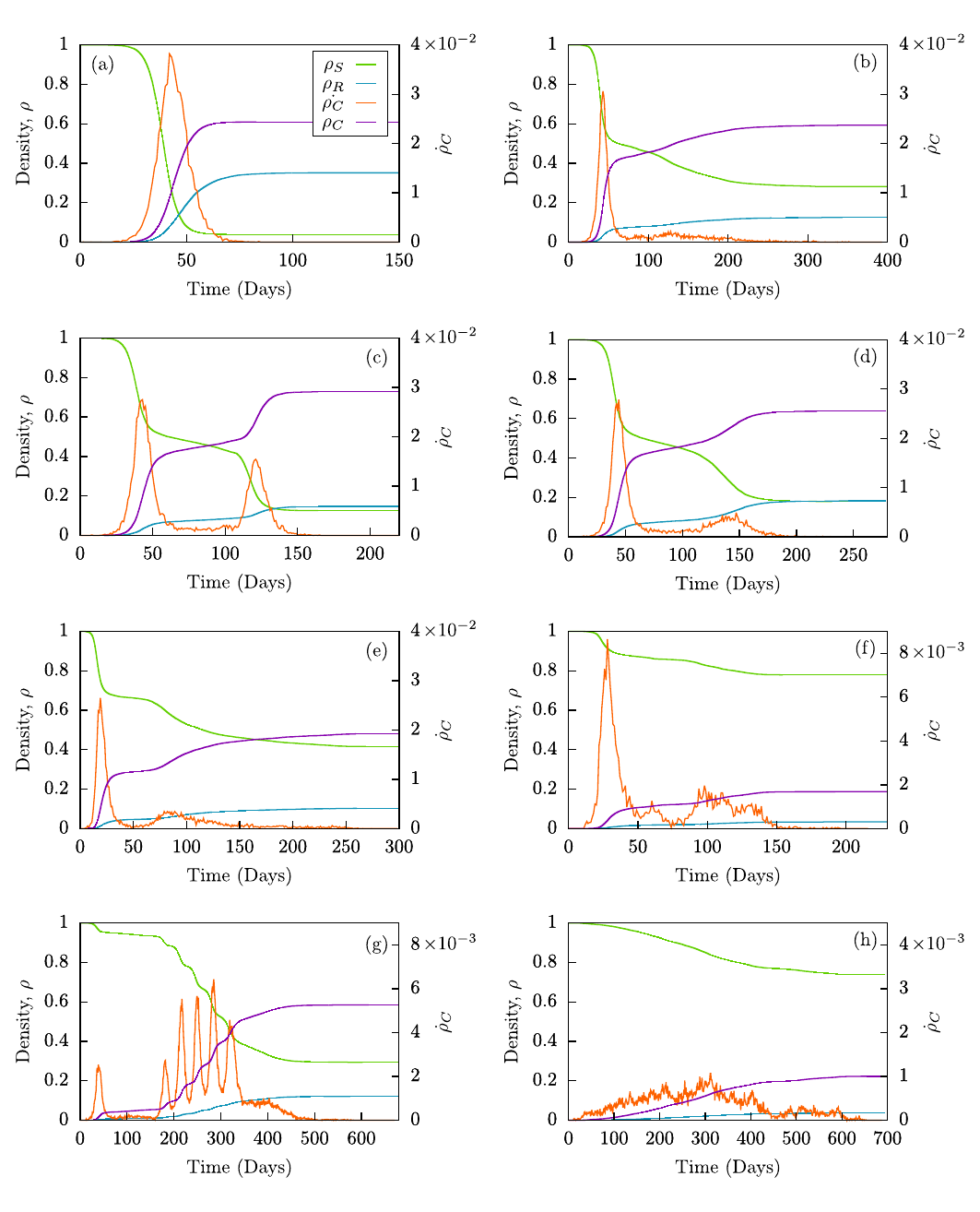}
\caption{
Plot of the densities $\rho_S(t)$, $\rho_R(t)$, $\dot{\rho}_C(t)$, and ${\rho}_C(t)$, where $\rho_C$ denotes $\rho_{R^*}+\rho_{I_a^\prime}+\rho_{I_s^\prime}$ as a function of $t$ for the Susceptible-Exposed-Infected-Recovered (SEIR) model. These represent the proportions of susceptible individuals, recovered individuals without noticing, newly confirmed cases, and accumulated confirmed cases, respectively. The rates are taken as $k_{1}=0.11$, $k_2=0.39$, $p_a=0.36$, $k_3=0.33$, $k_4=0.11$, and $k_5=0.08$. (a), Simulations are performed on ER random networks without the K-quarantine measures. System size $N=2.1\times 10^4$, the mean degree $\langle d \rangle=10$, and $k_6=0$ are set. (b), Similar plot to (a), but under the K-quarantine strategy $k_6=0.09$. (c), Similar plot to (b), but the rate $k_1$ changes suddenly at $t=108$ to $k_1=0.41$. This change is caused by a new type of coronavirus, GH clade~\cite{KCDC, Korber2020}. (d), Similar plot to (b), but the rate $k_6=0$ at $t=108$. This change is considered to occur because the quarantine system no longer functions owing to overloading. (e) and (f), Similar plots to (b), but simulations are performed on scale-free networks with degree exponent $\lambda=2.5$~\cite{Chung2002} and on an empirical social network~\cite{Lee2010}, respectively. For (f), $N=21403$ and $\langle d \rangle=7.8$. Owing to this smaller mean degree, the contagion rate is lower. (g) and (h), Similar plots to (b), but on modular networks. The network is composed of $N_m$ modules and each module contains $N_n$ nodes and has the mean degree of intra-module edge $\langle d_{\rm intra}\rangle=10$. Those modules are connected through $L_m$ inter-modular links. For (g), $N_m=10$, $N_n=10^4$, and $L_m=200$ are set. For (h), $N_m=10^3$, $N_n=10^2$, and $L_m=10^4$ are set.
\label{fig:densities}
}
\end{figure*}

\section{Reaction rates}
To explore the effect of the self-quarantine measure on the transmission of Covid-19, the rates $k_2-k_5$ and $p_a$ were estimated based on empirical data on Covid-19 provided by the Center for Disease Control (CDC) and KCDC. First, to find the rate $k_2$ and $p_a$, the time period between exposure and the onset of symptoms is used. This interval was estimated to be $6$ (mean) days~\cite{vespignani-global1, CDC-model, presymp2}.
The infected individual can transmit the disease 1$-$3 days prior to the onset of symptoms~\cite{presymp2, CDC-model}. Thus, the interval between exposure and becoming infectious is estimated to be $4$ (mean) days~\cite{vespignani-global2, realData1, realData2, realData3, incubation1, incubation2}. We take $k_2(1-p_a)=0.25$.
The resulting data show that the percentage of asymptomatic infections is estimated to be 15\%--40\%~\cite{asympRatio1,asympRatio2,asympRatio3}. We take $p_a=0.36$. Thus, $k_2\approx 0.39$ and $k_2 p_a\approx 0.14$.

In South Korea, a potential symptomatic infectious individual develop symptoms and then quarantined approximately in three days~\cite{presymp2, CDC-model}. Thus, $k_{3}\approx 0.33$ was set. Further, it takes approximately nine and 12 days for an asymptotic carrier and a confirmed infected individual, respectively, to recover~\cite{recover}. Thus, $k_{4}\approx 0.11$ and $k_{5}\approx 0.08$ were set. $\tau$ was taken to be 14 days.

The infection rate $k_1$ is estimated using the relation $R_0=k_1\langle d \rangle/k_3$. Using the rate $k_3=1/3$ and the mean degree $\langle d \rangle=10$, $k_1=0.11$ is obtained when $R_0=3.3$ is taken. Using these parameter values, it is observed that the simulation result fits the empirical data from the early stages of the Covid-19 outbreak in South Korea (March 2020) to the end of August. For the same outbreak, the value of $R_0$ directly measured from the empirical data is $R_0\approx 3.58$~\cite{korea_reproduction}.

Here, the K-quarantine model was simulated with fixed rates ($k_1-k_5$) and $p_a$, and a controllable quarantine rate ($k_6$) on several types of networks.
These included random networks (Fig.~\ref{fig:densities}(a)-(d)) scale-free networks (Fig.~\ref{fig:densities}(e)), an empirical social network (Fig.~\ref{fig:densities}(f)), and random networks with modules (Fig.~\ref{fig:densities}(g)-(h))~\cite{Watts2005,Colizza2007}.
It should be noted that all the rates are fixed throughout the epidemic spreading process unless otherwise specified. The proportions of nodes in each state are measured as a function of time in days.

\section{Temporal bahaviors of several quantities}
\begin{figure*}[ht]
	\centering
	\includegraphics[width=0.99\linewidth]{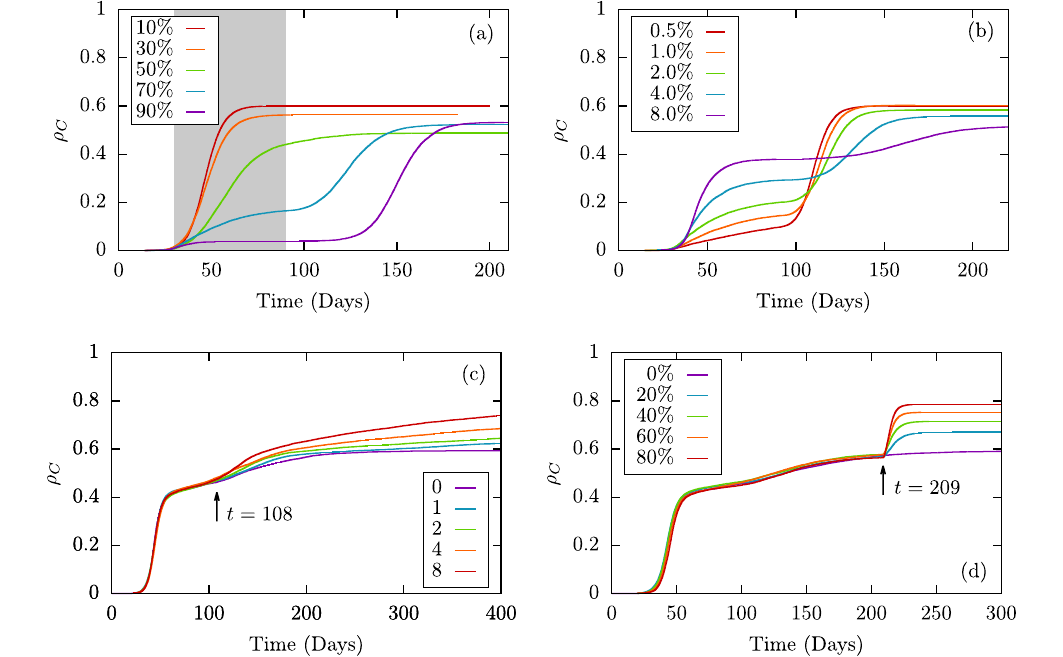}
	\caption{Similar plots to Fig.~\ref{fig:densities}(b). (a) But, at $t=30$, some fraction of links (indicated in legend) are artificially deleted. This change is considered to occur when a global lockdown is functioning. Depending on the  fraction, diverse temporal patterns of $\rho_C$ appear. (See the details in the main text.) (b), Similar plot to (a) but the lockdown is functioning when the fraction of accumulated confirmed cases reaches a threshold value given in legend. Then 70\% links are deleted at random. (c), Similar plot to Fig.~\ref{fig:densities}(b), but a small fraction of the nodes in state $S$ randomly selected everyday are forced to change their state to the latent state from the specified day. This is caused by the transmission of the disease by people from abroad. (d), Similar plot to Fig.~\ref{fig:densities}(b), but a large fraction of the nodes in state $S$ instantaneously change state to $L$ on a single occasion~\cite{Liu1986}. This change reflects the transmission of the disease by close contact among people participating in a large street demonstration.
	\label{fig:isolation}
	}
\end{figure*}

In Fig.~\ref{fig:densities}(a), the SEIR model~\cite{Kuznetsov1994,Li1999,Wang2020} is considered without any quarantine on random networks. Thus, $k_{6}=0$ was set. Initially, one node is assumed to be infected, while the other nodes are susceptible. The fractions $\rho_S(t)$, $\rho_R(t)$, $\dot{\rho}_C$, and $\rho_C$, are obtained, where $\rho_C=\rho_{R^*}+\rho_{I_a^\prime}+\rho_{I_s^\prime}$, and the dot represents the time derivative. $\dot{\rho}_C$ and $\rho_C$ represent the proportions of newly confirmed cases and the accumulated confirmed cases, respectively. The three densities are shown in Fig.~\ref{fig:densities}(a). The contagion spreads rapidly during the early stage and eventually reaches a steady state. As shown in Fig~\ref{fig:densities}(b) with $k_{6}=0.09$, when the quarantine system is functioning, the fraction $\rho_C$ initially increases rapidly, then slowly increases with some fluctuations, and finally reaches a steady state. Resurgent behavior is observed in ${\dot \rho}_C$. Further, it is noted that for the system with no quarantine strategy, the absorbing state of the infectious node completely disappears on reaching the 150th day, whereas for the K-quarantine system, it reaches the 400th day. The proportions of the accumulated confirmed cases for (a) and (b) are close; however, the proportion of remaining susceptible people is extremely small for (a), but it is more than 20\% for (b). On the other hand, the fraction of asymptomatic infected patients appears to be about 30\% for (a), but it is approximately 10\% for (b). This is because asymptomatic patients can be detected when they are in quarantine owing to the infection of their neighbors.

Fig.~\ref{fig:densities}(c) depicts the case in which the infection rate $k_1$ suddenly increases to $k_1=0.41$ at $t=108$, owing to the change of virus species from S or V to GH clade~\cite{KCDC, Korber2020}. There exists another significant peak of $\dot{\rho}_C$ around $t=130$, and the infection rate increases dramatically. Following this, the system reaches a steady state. The density of $\rho_C$ in the steady state increased by 22.72\% compared to that of case (b). However, no such dramatic change is observed in the empirical data. Fig.~\ref{fig:densities}(d) depicts the case in which the quarantine system is overloaded and does not act at a certain time (e.g., $t=108$). Then, $\dot \rho_C$ instantaneously exhibits resurgent behavior and $\rho_C$ rapidly increases and reaches a steady state, as in the SIR model.

Next, the K-quarantine model is simulated on a scale-free network in Fig.~\ref{fig:densities}(e) and on an empirical social network in Fig.~\ref{fig:densities}(f). It is thus concluded that overall, the contagion pattern is insensitive to network structure. However, the peaks of the daily confirmed case (f) is higher than those in Fig.~\ref{fig:densities}(b). For (f), the mean degree is smaller than that for Fig.~\ref{fig:densities}(a)-(e). Therefore, the proportion of accumulated confirmed cases in the steady state is considerably smaller than that in Fig.~\ref{fig:densities}(b). For (g)-(h), simulations are performed on modular networks~\cite{hierarchical1, hierarchical2}. The modular networks are composed of $N_m$ modules, each of which contains $N_n$ nodes. Thus, the total number of nodes in the system is $N_m N_n$. Nodes within each module are connected to each other randomly with mean degree $\langle d_{\rm intra}\rangle=10$. To make the modules connected, $\ell_m$ pairs of modules are selected randomly, each of the pairs are connected by $\ell_p$ links by selecting $\ell_p$ nodes from each module. Thus, the total number of inter-modular edges is $L_m=\ell_m \ell_p$. Specific those numbers are listed in the caption of Fig.~\ref{fig:densities}(g)-(h).

Fig.~\ref{fig:isolation}(a) and (b) depict the cases in which the system is lockdown for 60 days. The lockdown can be realized by either social distancing or restriction of transportation~\cite{reka, vespignani-global1, vespignani-global2}. In (a), the lockdown is implemented by deleting the fractions of links (indicated in the legend) randomly selected at the 30th day. After 60 days, those links are recovered. When the fractions are below 50\%, the lockdown effect is almost negligible. On the other hand, when the fraction is 90\%, then the epidemic spread is highly suppressed. In the intermediate range, a resurgent behavior appears. However, such behaviors fluctuate depending on the density of infectious nodes at the 30th day. Thus, in (b), we consider the case that the starting day of lockdown measure is determined by the fraction of accumulated confirmed cases, called lockdown threshold. Once the lockdown comes into force, 70\% of links are deleted and they are recovered after 60 days. Depending on the threshold value, the time of resurgent peak is determined. In short, while the lockdown measure during the 60 days is effective during some interval, the outbreak eventually occurs.

In Fig.~\ref{fig:isolation}(c), a small number of the nodes in state $S$ everyday change its state to $L$. This change is considered to occur when individuals from abroad become new sources of epidemic. Because in this case, no root is found explicitly and implicitly in the trail of disease transmission, the pattern of spread may somewhat differ from the previous patterns. In Fig.~\ref{fig:isolation}(d), a large fraction of the nodes in state $S$ instantaneously change state to $L$. This change is considered to occur by the transmission of disease among people participating in a large street demonstration at $t=209$ owing to their close contact and shouting.

\begin{figure}[!htb]
\centering
\includegraphics[width=0.99\linewidth]{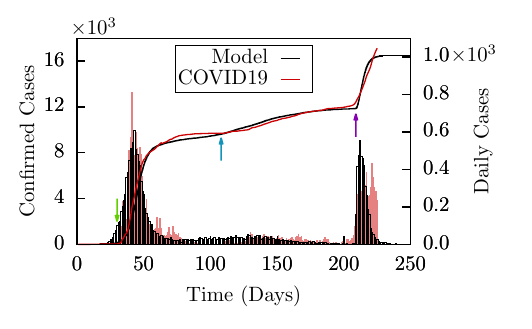}
\caption{(Red solid curve) Plot of the number of accumulated confirmed cases that occur in South Korea versus time in days. (Black dashed curve) Plot of the same quantity obtained from the K-quarantine model on an ER random network. The arrows indicate the dates of the three surges. With the choice of reaction rates $k_1-k_6$, the increasing behavior of the accumulated confirmed cases from the model is well-fitted to the empirical data up to $t=209$. Beyond this point, owing to an explosive epidemic contagion by a large number of street demonstrations, the theoretical curve no longer matches the empirical curve.
\label{fig:empirical}
}
\end{figure}

\begin{figure}[!htb]
\centering
\includegraphics[width=0.99\linewidth]{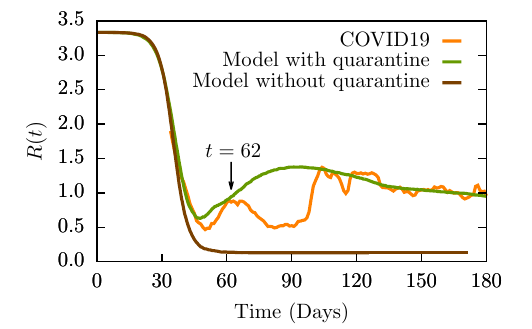}
\caption{Plots of three temporal reproduction numbers $R(t)$ estimated i) from the statistics of empirical daily confirmed cases~\cite{R_t} (orange), by simulations ii) with and iii) without the quarantine process (green and brown), respectively. Owing to the different methodologies of i) and ii), there exist some time delay between the two curves of $R(t)$. We shift the curve $R(t)$ of i) by 12 days to the left to overlap the two curves of i) and ii) in the early stage. For iii), $R(t)$ is obtained from simulation without the quarantine process. At March 21st ($t=62$), the Korean government increases somewhat the level of social distancing. As a result, the two curves of i) and ii) are affected. This enhanced prevention maintains until April 19th ($t=91$). 
\label{fig:Rt}
}
\end{figure}

In Fig.~\ref{fig:empirical}, the simulation results are compared with the empirical data of South Korea (accumulated as of September 9th, 2020). It is observed that the increasing behavior of the number of accumulated confirmed cases from the model during the early stage is well-fitted to the empirical data with the rates assumed herein. However, there is some difference during the intermediate stage, which may be due to the unexpected social event (a festival opening in a club) that was held shortly after reducing the level of social distancing. In the later region, the number of confirmed cases abruptly increases owing to the large demonstration on the main street near the city hall in Seoul. Among over 10,000 people participating in the demonstration, a non-negligible portion of them did not wear masks. Therefore, the disease transmission would be high. The model proposed herein cannot reproduce the output of such a large-scale perturbation. Instead, some portion (80\%) of the remaining susceptible nodes were changed to nodes in the latent state, under the assumption that those portions of people are infected in high-risk areas. With the passage of time, the surge decreases owing to the K-quarantine measures.

In Fig.~\ref{fig:Rt}, we plot three temporal reproduction numbers $R(t)$ estimated i) from the statistics of empirical daily confirmed cases provided in~\cite{R_t} (orange), by simulations ii) with and iii) without the quarantine measure (green and brown), respectively. For i), $R(t)$ is obtained as the ratio of the number of new infectious patients $I_t$ generated at time step $t$ to the total number of infectious patients during all precedent time steps, i.e., $\sum_{s=1}^{t-1}I_{t-s} w_s$, weighted with an infectivity function $w_s$. Moreover, the ratio is averaged over a time window of size $\tau$ ending at time $t$. Accordingly, the curve has little noise. For ii), $R(t)$ is obtained by the formula $k_1\langle d(t) \rangle/k_3$, where $k_1$ and $k_3$ are fixed,  and the mean number of susceptible neighbors of each infectious node at a given time $\langle d(t) \rangle$ is variable. Owing to the different methodologies, there exists some time delay between the two curves $R(t)$ of i) and ii). We shift the curve $R(t)$  of i) by 12 days to the left to make the two curves of i) and ii) overlap in the early stage. For iii), $R(t)$ is obtained from simulations without the quarantine process. While the two $R(t)$ curves are close to each other in the region $t > 105$, they are not in agreement with each other in the interval about $62 < t < 105$. This deviation may be caused by increasing the level of social distancing by the Korean government.

\section{Conclusion}
In summary, a network model was introduced to illustrate the spread of Covid-19 in South Korea under the K-quarantine model.
This model is essentially a SEIR model on networks; however, it also includes the process of disconnecting links around infected nodes. While these disconnections indicate the local isolation of infectious people, they do not necessitate global lockdown over the entire system.
It may be noted that social contact is not static but changes temporally. Thus, recognizing all individuals who were in contact with infectious individuals is a challenging task. Further, information on the spatial and temporal trajectories of infectious individuals is collected using diverse methods such as CCTV recordings, mobile phone data from local stations, and sending messages on mobile phones of ordinary people living in the given regions of the trajectories. This type of tracing requires a significant amount of human labor combined with advanced technology.

\begin{appendix}
\section{Simulation algorithm}
The probability that the state $X_{i}$ of a node $i$ changes the state $A$ from $B$ in unit time is called the transition rate $r_{X_{i}:A\to B}$.
For example, in the K-quarantine model, the probability that a node in the susceptible state $S$ moves to the latency state $L$ is as follows:
\begin{linenomath*}\begin{align}
	r_{X_{i};S\to L} = k_{1}\sum_{j}A_{ij}\delta(X_{j},L)
\end{align}\end{linenomath*}
where $A_{ij}$ is the adjacency matrix of the network, and $\delta(X_{j},L)$ is the Kronecker delta.

When the K-quarantine model is simulated through the discrete-time approach method, the state of each node is changed independently depending on the reactions in Fig.~\ref{fig:diagram}. On the other hand, the time interval $\tau$ is constant~\cite{Fennell2016}.
It is possible that node $i$ in state $A$ changes to $B_{1},\cdots, B_{M_{i}}$, and the transition rate of node $i$ is expressed as
\begin{linenomath*}\begin{align}
	r_{i} = \sum_{\alpha=1}^{M_{i}} r_{X_{i}:A\to B_{\alpha}}.
\end{align}\end{linenomath*}
Assume that the reactions follow the Poisson process. Then, for a given time interval $\tau$, the probabilities $p_{i}$ and $p_{X_{i}:A\to B_{\alpha}}$ that the state of the node $i$ changes to another state and to a specific state $B_{\alpha}$, respectively, are obtained as follows:
\begin{linenomath*}\begin{align}
	p_{i} = 1- e^{-r_{i}\tau},\quad p_{X_{i}:A\to B_{\alpha}} = \frac{ r_{X_{i}:A\to B_{\alpha}}}{r_{i}} p_{i},
\end{align}\end{linenomath*}
Therefore, a random number $u$ is chosen from the uniform random distribution following $(0:1]$.
If $u>p_{i}$, the state of node $i$ is not changed; else, the state of node $i$ may be changed to $B_{\beta}$ as follows:
\begin{linenomath*}\begin{align}
	\sum_{\alpha=k}^{\beta-1} r_{X_{i}:A\to B_{\alpha}}<u\leq \sum_{\alpha=1}^{\beta} r_{X_{i}:A\to B_{\alpha}}.
\end{align}\end{linenomath*}
Following the updation of the states of all nodes in parallel, this increases the time by $\tau$.

The Gillespie algorithm~\cite{Gillespie1977} was also employed for the numerical simulation of the K-quarantine model.
Because this algorithm adjusts the time interval $\tau$ according to the transition probability, it is widely used to simulate stochastic epidemic models in real time~\cite{Ferreira2012,Vestergaard2015}.
The cumulative transition rate $r$ is the probability that at least one reaction occurs per unit time,
\begin{linenomath*}\begin{align}
	r = \sum_{i=1}^{N} r_{i}.
\end{align}\end{linenomath*}
It is well known that the time interval $\tau$ in which at least one reaction occurs follows an exponential distribution with a mean of $1/r$.
\begin{linenomath*}\begin{align}
	P(\tau) = r e^{-r \tau}
\end{align}\end{linenomath*}
Two random numbers $u_{1}$ and $u_{2}$ were chosen from the uniform random distribution following $(0:1]$, respectively.
Then, the time interval $\tau= \frac{1}{r}\ln\frac{1}{u_{1}}$ was found, and the node $j$ and state $B_{\beta}$ were obtained as follows:
\begin{linenomath*}\begin{align}
	\sum_{i=1}^{j}\sum_{\alpha =1}^{\beta-1} r_{X_{i}:A\to B_{\alpha}} < ru_{2}\leq \sum_{i=1}^{j}\sum_{i=\alpha}^{\beta}r_{X_{i}:A\to B_{\alpha}}.
\end{align}\end{linenomath*}
Following the change of the state of node $j$ to state $B_{\beta}$, the time increases by $\tau$.

\begin{acknowledgments}
	This research was supported by the NRF, Grant No.~NRF-2014R1A3A2069005 (BK). BK thanks Prof. H. B. Kim MD for helpful discussions.
\end{acknowledgments}


\end{appendix}

\bibliography{ref}

\end{document}